\documentclass[
preprint,
showpacs,
preprintnumbers,
nofootinbib,
bibnotes,
amsmath,amssymb,
aps,
prl,
]{revtex4-1}

\usepackage[pdftex]{graphicx}
\usepackage{dcolumn}% Align table columns on decimal point
\usepackage{bm}% bold math
\usepackage{color}
\usepackage{multirow}
\usepackage{mathrsfs}
\usepackage{amssymb}
\usepackage{amsmath}

\begin{document}

%\preprint{\large{\textit{Draft: Version 3}}}

\title{Phenomenology of $n$-${\bar n}$ oscillations revisited}

\author{S.\ Gardner}
\email{\texttt{gardner@pa.uky.edu}}
\author{E.\ Jafari}
%\email{\texttt{jafari@pa.uky.edu}}
\affiliation{Department of Physics and Astronomy, University of Kentucky,
Lexington, Kentucky 40506-0055 USA}

%\date{\today}
%%%%%%%%%%%%%%%%%%%%%%%%%%%%%%%%%%%%%%%%%%%%%%%%%%%%%%%%%%%%%%%%%%%%%%%%%%%%%

\begin{abstract}
We revisit the phenomenology of $n$-${\bar n}$ oscillations 
in the presence of external magnetic fields, highlighting the role of spin. 
We show, contrary to long-held belief, that 
the $n$-${\bar n}$ transition rate need not be suppressed, 
opening new opportunities for its empirical study. 
\end{abstract}

\maketitle

%%%%%%%%%%%%%%%%%%%%%%%%%%%%%%%%%%%%%%%%%%%%%%%%%%%%%%%%%%%%%%%%%%%%%%%%%%%%%%%

%\section{Introduction}
%\label{sec:introduction}
%%%%%%%%%%%%%%%%%%%%%%%%%%%%%%%%%%%%%%%%%%%%%%%%%%%%%%%%%%%%%%%%%%%%%%%%%
%%%%%%

{ \em 1. Introduction.} Searches for 
processes that violate standard model (SM) symmetries 
are of particular interest because their discovery 
would serve as unequivocal evidence for 
dynamics beyond the SM.
The gauge symmetry and known 
particle content of the SM implies that its Lagrangian conserves
baryon number ${\cal B}$ and lepton number ${\cal L}$, 
though it is the combination 
${\cal B} - {\cal L}$ that survives at the quantum level. 
Thus the observation of neutron-antineutron 
($n$-${\bar n}$) oscillations, a $|\Delta {\cal B}|=2$ process, 
would show that 
${\cal B}- {\cal L}$ 
symmetry is broken and ergo that 
dynamics beyond the SM exists. The current constraints on 
$|{\cal B}|=1$ operators from the non-observation of nucleon decay 
are severe, with the strongest limits
coming from searches for 
proton decay to final states that respect ${\cal B}- {\cal L}$ symmetry, 
such as $p\to e^+ \pi^0$, for which the partial half-life exceeds 
$8.2 \times 10^{33}\,{\rm years}$ at 90\% C.L.~\cite{Beringer:1900zz}.
Although particular $|\Delta {\cal B}|=1$ operators, such as those that 
mediate $n\to e^- \pi^+$, e.g., can 
also give rise to $n$-${\bar n}$ oscillations, 
Mohapatra and others have emphasized that the origin of 
nucleon decay and $n$-${\bar n}$ oscillations 
can be completely different~\cite{Mohapatra:1980qe,Babu:2001qr,Mohapatra:2009wp,Ajaib:2009fq,Babu:2012vc,Arnold:2012sd,Babu:2013yca}. 
Recently, moreover, simple models 
that give rise to $n$-${\bar n}$ oscillations 
but not nucleon decay have been enumerated~\cite{Arnold:2012sd}.

The seminal papers on free $n$-${\bar n}$ oscillations have 
employed a $2\times 2$ effective Hamiltonian
matrix~\cite{Mohapatra:1980de,Cowsik:1980np}, familiar from the 
analysis of meson mixing~\cite{Bigi:2000yz}, though this choice 
explicitly suppresses the role of spin --- unlike neutral mesons and neutrinos, 
the neutron and antineutron each have a significant magnetic moment.  
We note the neutron and antineutron 
are themselves distinguished by the sign of the lepton charge in 
semileptonic decay, and their respective interactions with 
atomic nuclei are strikingly different as well~\cite{Dover:1982wv,Dover:1985hk}. 
The $n$-${\bar n}$ system thus has four 
physical degrees of freedom because the spin projection of a neutron or an 
antineutron can either be 
parallel or antiparallel to a quantization axis. 
In this paper we develop a suitable 
$4\times 4$ effective Hamiltonian framework for its study. 
Since previous studies of $n$-${\bar n}$ oscillations have 
been realized in the context of a $2\times 2$ effective Hamiltonian
matrix, we discuss this framework 
before turning to our generalization. 
The neutron magnetic moment is empirically well-known, yielding
an interaction with an external magnetic field $\mathbf{B}$
of form $-\mu_n \mathbf{S}_n \cdot \mathbf{B}/S_n$, 
where $\mu_n$ is the magnitude of the magnetic moment and 
$\mathbf{S}_n$ is the neutron spin. 
Supposing the neutron spin to be in the 
direction of the applied
$\mathbf{B}$-field 
and employing \textbf{CPT} invariance, 
the mass matrix ${\cal M}$ takes the form~\cite{Mohapatra:1980de}
\begin{equation}
{\cal M}=\left(\begin{array}{cc}
M_n - \mu_n B  & \delta \\
\delta  & M_n + \mu_n B 
\end{array}\right)\,,
\label{eq2by2}
\end{equation}
where \textbf{CPT} invariance guarantees not only that
the neutron and antineutron masses are equal but also that the projections of
the neutron and antineutron magnetic moments on $\mathbf{B}$ are equal in
magnitude and of opposite sign. We work in units $\hbar=c=1$ 
and ignore the finite neutron and antineutron lifetimes throughout. 
Diagonalizing ${\cal M}$ yields the mass 
eigenstates $|u_i\rangle$, namely, 
\begin{eqnarray}
|u_{1}\rangle &=& \cos\theta |n\rangle+ \sin\theta |\overline{n}\rangle \,,
\nonumber\\
|u_{2}\rangle &=& -\sin\theta |n\rangle + \cos\theta |\overline{n}\rangle \,. 
\end{eqnarray}
Since the energy scale $\mu_n B$ naturally 
dwarfs that of $\delta $, 
we note that the eigenvalue difference is 
$\Delta E \simeq 2 \mu_n B$ and that $\theta$ is small: 
$\theta \simeq \delta /\Delta E$. 
The $n$-${\bar n}$ transition probability becomes~\cite{Kronfeld:2013uoa}
\begin{equation}
P_{\overline{n}}(t)\simeq 2 \theta^{2} 
\left[1 - \cos \left(\Delta E t\right) \right] \,. 
\end{equation}
This result can be considered 
in two different limits: either
(a) $\Delta E t\gg1$ or (b) $\Delta E t\ll 1$. 
In case (a) the second term oscillates to zero, yielding 
$P_{\overline{n}}(t)\simeq 2 (\delta /\Delta E)^2$
whereas in case (b), 
\begin{equation}
P_{\overline{n}}(t) \simeq  
\left(\frac{\delta }{\Delta E}\right)^2 (\Delta E t)^2 = (\delta  t)^2 \,.
\label{usual}
\end{equation}
Evidently unless $t\ll 1/\Delta E$, the energy splitting of
the neutron and antineutron in a magnetic field ``quenches''
the appearance of $n$-${\bar n}$ oscillations. Thus the 
strategy in past and proposed searches for $n$-${\bar n}$ oscillations has been
to minimize the magnetic field~\cite{BaldoCeolin:1994jz,Kronfeld:2013uoa,ESS}, 
so that $t \ll 1/\Delta E$, as well as to maintain 
a vacuum in the neutron flight volume~\cite{Cowsik:1980np}, 
so that the neutrons are quasifree 
over the neutron observation time $t$. 

Motivated by the realization  
that a neutron and an antineutron of opposite spin projection have the same energy
in a magnetic field, we consider the spin dependence of 
$n$-${\bar n}$ oscillations explicitly and thus develop a 
$4\times 4$ effective Hamiltonian 
framework for its analysis. Spin dependence can arise from effects either within 
or beyond the SM. As long known from the theory of magnetic resonance, applied magnetic fields can 
mitigate, or even remove, the energy splitting of spin states in a static
magnetic field, note, e.g., Ref.~\cite{PhysRev.51.652,CTDL}. 
In this paper we show that such SM effects can remove the magnetic field ``quenching'' 
noted in the usual $2\times 2$ Hamiltonian framework and yield new experimental possibilities
for the study of $n$-${\bar n}$ mixing. 
It is also possible to have new, 
spin-dependent ${\cal B}-{\cal L}$ violating operators, yielding a 
``new physics'' mechanism to evade the magnetic field quenching we have noted. 
Although 
we consider both of these distinct possibilities in this paper, our primary
focus is the role of spin-dependent SM effects in mediating $n$-${\bar n}$ 
oscillations. 

{ \em 2. Effective Hamiltonian for $n$-${\bar n}$ transitions with spin.} 
To realize the most general form of a low-energy, 
phenomenological Hamiltonian for 
$n$-${\bar n}$ oscillations with spin, we develop
a mass matrix ${\cal M}$ to this purpose. Its entries ${\cal M}_{ij}$ 
with $i,j=1,\dots 4$ correspond to bras and kets containing 
$n(\mathbf{p},+)$, ${\bar n}(\mathbf{p},+)$, 
$n(\mathbf{p},-)$, and ${\bar n}(\mathbf{p},-)$, respectively, with 
$+\, (-)$ denoting a spin-up (down) state, relative
to a quantization axis $\mathbf{z}$. 
We impose the constraint of Hermiticity, as well as those of 
charge-conjugation--parity (\textbf{CP}) and 
time-reversal (\textbf{T})  
invariance, on the resulting mass matrix, to determine its model-independent 
form under these assumptions. 

We can implement the discrete symmetry transformations in 
relativistic quantum field theory and 
translate them to quantum mechanics by noting~\cite{Bigi:2000yz} 
\begin{equation}
\mathbf{b}^{\dagger}(\mathbf{p},s) | 0 \rangle = | n(\mathbf{p},s ) \rangle 
\quad; \quad
\mathbf{d}^{\dagger}(\mathbf{p},s) | 0 \rangle = | {\bar n}(\mathbf{p},s ) \rangle\,,
\end{equation}
where 
$\mathbf{b} [\mathbf{b}^{\dagger}] (\mathbf{p},s)$ and 
$\mathbf{d} [\mathbf{d}^{\dagger}] (\mathbf{p},s)$ 
denote annihilation {[}creation{]} operators for neutrons {[}antineutrons{]}
of momentum $\mathbf{p}$ and spin projection $s$, for which
$s=\pm 1\equiv \pm$ with respect to the quantization axis $\mathbf{z}$. 
We determine the transformation properties
of these operators under \textbf{CP} and \textbf{T} as follows. 
We work in the Dirac-Pauli representation for the $\gamma^\mu$ matrices
and note that the Dirac field operator $\psi(x)$ 
has a plane-wave expansion of form 
\begin{equation}
\psi(x) = \int \frac{d^3 \mathbf{p}}{(2\pi)^{3/2} \sqrt{2 E}}
\sum_{s=\pm} \left\{ b(\mathbf{p}, s) u(\mathbf{p}, s) e^{-ip\cdot x}
+ d^\dagger(\mathbf{p}, s) v(\mathbf{p}, s) e^{ip\cdot x}
\right\} \,,
\label{free} 
\end{equation}
with spinors defined as 
\begin{equation}
u(\mathbf{p}, s) = {\cal N}
\left(\begin{array}{c}
\chi^{(s)} \\
\frac{\mathbf{\sigma}\cdot \mathbf{p}}{E + M} \chi^{(s)} \\
\end{array}\right) 
\quad ; \quad
v(\mathbf{p}, s) = {\cal N}
\left(\begin{array}{c}
\frac{\mathbf{\sigma}\cdot \mathbf{p}}{E + M} \chi^{\prime\, (s)} \\
\chi^{\prime (s)} \\
\end{array}\right) \,,
\end{equation} 
noting $\chi^{\prime\, (s)} = -i \sigma^2 \chi^{(s)}$, 
$\chi^{+} = \left( \stackrel{1}{{}_0} \right)$,  
$\chi^{-} = \left(\stackrel{0}{{}_1} \right)$, and ${\cal N}=\sqrt{E + M}$.
This yields 
\begin{equation}
\mathbf{CP}\,\mathbf{b}(\mathbf{p},s)\,(\mathbf{CP})^{\dagger}
=\mathbf{d}(-\mathbf{p},s) \quad ; \quad 
\mathbf{CP}\,\mathbf{d}(\mathbf{p},s)\,(\mathbf{CP})^{\dagger}
=- \mathbf{b}(-\mathbf{p},s) 
\label{CPtrans}
\end{equation}
and 
\begin{equation}
\mathbf{T}\,\mathbf{b}(\mathbf{p},s)\,(\mathbf{T})^{-1}
= s \mathbf{b}(-\mathbf{p},-s) \quad ; \quad 
\mathbf{T}\,\mathbf{d}(\mathbf{p},s)\,(\mathbf{T})^{-1}
= s \mathbf{d}(-\mathbf{p},-s) 
\label{Ttrans}
\end{equation}
for the transformation properties under 
$\mathbf{CP}$ and $\mathbf{T}$, 
respectively{\footnote{These results differ from those in 
Ref.~\cite{Bigi:2000yz} because that work uses a different 
choice of antiparticle spinor.
}}. 
In what follows we assume 
that the ground (vacuum) state
remains invariant under \textbf{CP} and \textbf{T}: 
$\mathbf{CP}|0\rangle=|0\rangle$ and $\mathbf{T}|0\rangle=|0\rangle\,$. 

Under an assumption of \textbf{CP} and \textbf{T} invariance 
relationships between 
the matrix elements of ${\cal M}$ follow. 
For example, under 
\textbf{CPT} invariance we have
\begin{equation}
\langle n(\mathbf{p}, s_1) | H | n(\mathbf{p}, s_2) \rangle 
= s_1 s_2
\langle 
{\bar n}(\mathbf{p}, -s_2) | H | {\bar n}(\mathbf{p}, -s_1) 
\rangle  \,,  
\end{equation} 
noting $H$ is the Hamiltonian and 
\textbf{T} is an anti-unitary operator. 
Thus under \textbf{CPT}
and Hermiticity we find ${\cal M}$ has ten parameters, and it is of
form 
\begin{equation}
\left(\begin{array}{cccc}
A_{1} & \delta & M_{1} & \varepsilon_{1}\\
\delta^{\ast} & A_{2} & \varepsilon_{2} & -M_{1} \\
M_{1}^{\ast} & \varepsilon_{2}^{\ast} & A_{2} & -\delta \\
\varepsilon_{1}^{\ast} & -M_{1}^{\ast} & -\delta^{\ast} & A_{1}
\end{array}\right) 
\,,
\label{Mform}
\end{equation}
where $A_1$ and $A_2$
are real constants. Under \textbf{CP} invariance we have, e.g.:
$\langle n(\mathbf{p}, s_1) | H | n(\mathbf{p}, s_2) \rangle 
= 
\langle {\bar n}(-\mathbf{p}, s_1) | H | {\bar n}(-\mathbf{p}, s_2) \rangle$,
yielding relationships between $M_{ij}$ 
in the low-energy limit, i.e., as $|\mathbf{p}|\to 0$. 
Thus under Hermiticity and \textbf{CP} and \textbf{CPT} invariance we
have in this case 
\begin{equation}
\left(\begin{array}{cccc}
A_{1} & i \delta & 0 & \varepsilon_{1}\\
-i \delta & A_{1} & -\varepsilon_{1} & 0 \\
0 & -\varepsilon_{1}^{\ast} & A_{1} & -i \delta \\
\varepsilon_{1}^{\ast} & 0 & i \delta & A_{1}
\end{array}\right) 
\,,
\label{MformCP}
\end{equation}
where both $A_1$ and $\delta$ are real --- and only four parameters
suffice to characterize the mass matrix. 
In Eq.~(\ref{MformCP}), 
two distinct $n$-${\bar n}$ transition operators
appear: $\delta$ that describes the transition between states of 
the same spin, $n(s) \leftrightarrow {\bar n}(s)$ and $\varepsilon_1$ that 
describes the transition between states of opposite spin, 
$n(s) \leftrightarrow {\bar n}(-s)$. Note that since 
the neutron and antineutron are of 
opposite intrinsic parity, we have under
\textbf{CP}, 
$\langle n(\mathbf{p}, s_1) | H | {\bar n}(\mathbf{p}, s_2) \rangle 
=- \langle {\bar n}(-\mathbf{p}, s_1) | H | n(-\mathbf{p}, s_2) \rangle$, 
yielding, e.g., terms in $\pm i \delta$. If, rather, the relevant
piece of $H$ is odd under \textbf{CP}, the $\delta$ terms
become real, as chosen in Eq.~(\ref{eq2by2}). 
Previous analyses~\cite{Kronfeld:2013uoa} 
have only considered
the possibility of $n(s) \leftrightarrow {\bar n}(s)$. 
We will show that the
second process can occur through the application of magnetic fields, both
within and beyond the SM. The parameters $\delta$ and $\varepsilon_1$, however,
characterize $n$-${\bar n}$ mixing {\it en vacuo}. 
Since we have chosen the antiparticle spinors in a manner
consistent with Dirac hole theory, the underlying 
two-component spinor of a particle with spin $s$ has the same orientation 
as that of an antiparticle with spin $-s$; in the presence of baryon-number
violation it would seem that both pathways could occur. 
Indeed there are two Lorentz-invariant, leading-mass-dimension $n$-${\bar n}$ operators: 
$i n^T C n$ and $n^T \gamma_5 C n$, where $C=i\gamma^2 \gamma^0$ and $T$ denotes 
transpose.
The latter operator, $n^T \gamma_5 C n$, can potentially yield a spin flip. 
The leading-mass-dimension 
operators that yield $n$-${\bar n}$ transitions have been analyzed
in QCD~\cite{Rao:1982gt,Caswell:1982qs}, and they
entrain both possibilities at the quark level. 
Our detailed analysis of their $n$-${\bar n}$ matrix elements 
reveals, however, that 
$n(s) \leftrightarrow {\bar n}(-s)$  does not occur (at zero momentum transfer)~\cite{svgejxy}, 
as one might
expect from angular momentum conservation. 
Indeed only the $n(s) \to {\bar n}(s)$ transition occurs for a free neutron in vacuum. 
The associated $n$-${\bar n}$ matrix elements 
have been computed in 
models~\cite{Rao:1982gt,Fajfer:1983ja} and in lattice QCD~\cite{Buchoff:2012bm}. 
Thus we set $\varepsilon_1=0$ henceforth, 
though such could be nonzero in the presence of a hidden U(1) 
sector with 
a ``dark photon'' and an associated magnetic field  $\mathbf{B}_{\rm hidden}$.
Returning to the operators $i n^T C n$ 
and $n^T \gamma_5 C n$, the first is 
\textbf{CP} odd, whereas the second is \textbf{CP} even --- and both are 
\textbf{CPT} invariant. We assumed the second case in 
determining Eq.~(\ref{MformCP}), and this will prove useful in what
follows. However, since $n$-${\bar n}$ transitions in the absence of a magnetic
field are, in effect, mediated by $i n^T C n$, we use 
\begin{equation}
\left(\begin{array}{cccc}
A_{1} & \delta & 0 & 0 \\
\delta & A_{1} & 0 & 0 \\
0 & 0 & A_{1} & - \delta \\
0 & 0 & - \delta & A_{1}
\end{array}\right) 
\,,
\label{MformCPfinal}
\end{equation}
with $\delta$ real 
for our Hamiltonian matrix in this case. 

These parametrizations also allow us to generalize our 
effective Hamiltonian framework to include external magnetic fields. 
For example, the interaction of an electrically neutral particle
with an electromagnetic field is characterized at low energies by 
$-\mathbf{\mu}\cdot \mathbf{B}$ 
if \textbf{T} and \textbf{P} are not broken; 
this comes from the nonrelativistic limit of 
$\bar\psi \sigma^{\mu\nu} \psi F_{\mu \nu}$, where 
$F_{\mu \nu}\equiv \partial_\mu A_\nu - \partial_\nu A_\mu$ is the
usual electromagnetic field strength tensor. Under \textbf{CP} or \textbf{T} 
the fermion bilinear $\bar\psi \sigma^{\mu\nu} \psi$ transforms to 
$-\bar\psi \sigma_{\mu\nu} \psi$, and
$F_{\mu\nu}$ transforms to $-F^{\mu \nu}$. 
Thus 
their scalar product is itself both \textbf{CP} and \textbf{T} invariant. 
However, the explicit \textbf{CPT} and \textbf{CP} constraints we have
investigated operate on the fermion and antifermion degrees of freedom
only; the terms in $H$ 
resulting from the 
overall minus sign 
associated with 
$F_{\mu \nu}$ 
under \textbf{CP} 
are revealed by comparing the parametrizations under Hermiticity and 
\textbf{CPT} with and without a
\textbf{CP} constraint, 
Eqs.~(\ref{Mform}) and (\ref{MformCP}). We can also combine 
magnetic-field interactions with $n$-${\bar n}$ oscillations through
the operator $\psi^T \sigma^{\mu\nu} C \psi F_{\mu\nu}$ and its Hermitian conjugate; 
this operator is even under \textbf{CP} and \textbf{T}. 
Thus through these comparisons we 
see how $F_{\mu\nu}$ terms, i.e., those with external magnetic fields, can enter
both within and beyond the SM. We now turn to concrete expressions for these terms. 

{ \em 3. Effective Hamiltonian for $n$-${\bar n}$ transitions in external 
magnetic fields.} 
The operator $\psi^T \sigma^{\mu\nu} C \psi F_{\mu\nu}$ and its Hermitian conjugate 
yield $n\to {\bar n}$ and ${\bar n}\to n$ transitions, respectively. 
Computing these matrix elements using the free Dirac field operator of Eq.~(\ref{free}) yields 
\begin{equation}
\langle {\bar n} (\mathbf{0}, s') | 
\psi^T \sigma^{\mu\nu} C \psi F_{\mu\nu} | n(\mathbf{0}, s) \rangle 
= - \chi^{\prime\, (s')\, \dagger}  2\mathbf{\sigma}\cdot \mathbf{B} \chi^{\prime\, (s)} 
- \chi^{(s)\, \dagger}  2\mathbf{\sigma}\cdot \mathbf{B} \chi^{(s)} \,,
\label{tran1}
\end{equation}
where we recall $\chi^{\prime\, (s)} = -i\sigma^2 \chi^{(s)}$, 
and
\begin{equation}
\langle {n} (\mathbf{0}, s') | -\psi^{\ast\,T} C (\sigma^{\mu\nu})^\dagger \psi^\ast F_{\mu\nu} 
| {\bar n}(\mathbf{0}, s) \rangle 
= - \chi^{\prime\, (s')\, \dagger}  2\mathbf{\sigma}\cdot \mathbf{B} \chi^{\prime\, (s)} 
- \chi^{(s)\, \dagger}  2\mathbf{\sigma}\cdot \mathbf{B} \chi^{(s)} \,. 
\label{tran2}
\end{equation}
Although these expressions vanish for elementary fermions, we 
note that since both $n$ and $\bar n$ possess 
anomalous magnetic moments compositeness could make 
these matrix elements nonzero 
if operators of form $\psi^T C \psi$ exist. 
We leave a detailed
study to a subsequent publication~\cite{svgejxy}. 
Nevertheless, 
these expressions correspond to nonrelativistic operators containing 
$n$-$\bar{n}$ transition magnetic moments. 
Thus we suppose the $n$ and $\bar{n}$ 
interactions in the presence of external magnetic fields, 
under \textbf{CPT} invariance, 
to be of form 
\begin{equation}
H_{B}=-\mu_{n}\frac{\mathbf{S}_n}{S_n}\cdot\mathbf{B} + 
\mu_{n}\frac{\mathbf{S}_{\bar n}}{S_{\bar n}}\cdot\mathbf{B}
- \mu_{n \bar{n}}^\ast \frac{\mathbf{S}_{{\bar n} n}}{S_{{\bar n} n}} \cdot \mathbf{B}
- \mu_{n \bar{n}} \frac{\mathbf{S}_{n\bar n}}{S_{n\bar n}} \cdot \mathbf{B}
\,,
\label{HamB}
\end{equation}
where $\mu_n$ is the neutron magnetic moment, the first two terms being 
the usual neutron and antineutron interactions in a magnetic field, 
and $\mathbf{\mu}_{n\bar{n}}$ is the $n$-$\bar{n}$ 
transition magnetic moment. The last two terms correspond to Eqs.~(\ref{tran1}) and 
(\ref{tran2}), respectively. 
The spin operators each act in a $2\times 2$ subspace. 
With $(S_n)_{i,j}$ such that $(i,j) \in (n(+), n(-))$, we choose
$(S_{\bar n})_{i,j}$ with $(i,j) \in ({\bar n(+)}, {\bar n}(-))$, 
as well as $(S_{n \bar n})_{ij}$ and $(S_{{\bar n} n})_{ji}$ with
$i\in n(+), n(-)$ and $j\in {\bar n}(+), {\bar n}(-)$. 
Within a given subspace, we compute 
$\mathbf{S}\cdot \mathbf{B}/S = \mathbf{\sigma}\cdot \mathbf{B}$. 
We also suppose that magnetic fields both longitudinal and transverse to the
quantization axis exist, and we introduce 
$\mathbf{B}_0=B_0\hat{\mathbf{z}}$ and $\mathbf{B}_1=B_1\hat{\mathbf{x}}$, 
respectively. 
Defining $\omega_0  \equiv -\mu_{n} B_0$, $\omega_1  \equiv -\mu_{n} B_1$, 
$\delta_0  \equiv -\mu_{n\bar n} B_0$, 
$\delta_1  \equiv -\mu_{n\bar n} B_1$, and 
employing the usual Pauli matrices, 
we find that the matrix 
${\cal H}_B$ corresponding to Eq.~(\ref{HamB}) is
\begin{equation} 
{\cal H}_B = 
\left(\begin{array}{cccc}
\omega_0 & \delta_0 
& \omega_1  & 
\delta_1
 \\
\delta_0^\ast   &  -\omega_0 & 
\delta_1^\ast  & -\omega_1  \\
\omega_1  & \delta_1 
 &  -\omega_0 & -\delta_0  
\\
\delta_1^\ast
& -\omega_1  & -\delta_0^\ast  &  \omega_0
\end{array}\right) 
\,,
\label{MformB}
\end{equation}
a form consistent with the comparison of 
Eq.~(\ref{Mform}) and Eq.~(\ref{MformCP}). 
Moreover, we see that \textbf{CPT} invariance guarantees that 
a neutron and an antineutron of opposite spin in vacuum are always
degenerate irrespective of the size of the magnetic field: 
the presence of external magnetic fields cannot quench transitions
between these states. 

Additional constraints on the form factors follow because in the
presence of 
$n$-${\bar n}$ oscillations 
the weak interaction eigenstates 
can be expressed in terms of Majorana states. 
A Majorana state $|\Psi_M\rangle$ transforms into itself 
under \textbf{C}, up to a global phase. Since 
$\mathbf{C}\,\mathbf{b}(\mathbf{p},s)\,\mathbf{C}^{\dagger}
= \mathbf{d}(\mathbf{p},s)$, 
\begin{equation}
| \Psi_M^\pm  (\mathbf{p}, s) \rangle = \frac{1}{\sqrt{2}}
\left( 
 | {\bar n} (\mathbf{p}, s) \rangle \pm  | n (\mathbf{p}, s) \rangle 
\right) \,.
\label{Major}
\end{equation}
As we have noted, the neutron and antineutron are 
distinguished by the sign of the lepton charge upon semileptonic decay, 
so that the Majorana basis has four degrees of
freedom. 
There are no $\gamma^\mu$, $\sigma^{\mu\nu}$, or
$\sigma^{\mu\nu}\gamma_5$ form factors associated with a
Majorana state~\cite{Nieves:1981zt,Schechter:1981hw,Kayser:1982br,Shrock:1982sc,Li:1981um,Davidson:2005cs}; 
thus the constraint
$\langle \Psi_M^\pm  (\mathbf{p}, s') \rangle | H_B |\Psi_M^\pm  (\mathbf{p}, s) \rangle =0$ 
or, equivalently, $\eta^T {\cal H}_B \eta=0$, where $\eta= \{a,a,b,b\}$ 
and $a$ and $b$ are 
arbitrary constants, yields ${\rm Re} (\delta_0)=0$ and ${\rm Re} (\delta_1)=0$. 
With these supplemental constraints, Eq.~(\ref{MformB}) becomes 
\begin{equation} 
{\cal H}_B = 
\left(\begin{array}{cccc}
\omega_0 & i\delta_0 
& \omega_1  & 
i\delta_1
 \\
-i\delta_0   &  -\omega_0 & 
-i \delta_1  & -\omega_1  \\
\omega_1  & i \delta_1 
 &  -\omega_0 & -i \delta_0  
\\
- i \delta_1
& -\omega_1  & i\delta_0  &  \omega_0
\end{array}\right) 
\,,
\label{MBfinal}
\end{equation}
where $\delta_0$ and $\delta_1$ are real constants. 
This bears comparison to studies of
resonant spin-flavor neutrino precession in matter, 
such as in the Sun~\cite{Okun:1986na,Okun:1986hi,Lim:1987tk}, though 
the neutrino transition magnetic moment in that work is associated
with the transverse magnetic field and is 
flavor-changing.  
The final Hamiltonian matrix ${\cal M}$ for low-energy, $n$-${\bar n}$ oscillations
in applied magnetic fields thus takes the form
\begin{equation} 
{\cal H} = \left(\begin{array}{cccc}
M + \omega_0 & (\delta + i\delta_0) 
& \omega_1  & 
i\delta_1
 \\
(\delta - i \delta_0)   &  M  -\omega_0 & 
-i \delta_1  & -\omega_1  \\
\omega_1  & i \delta_1 
 & M -\omega_0 & - (\delta + i \delta_0) 
\\
- i \delta_1
& -\omega_1  & -(\delta - i \delta_0)  & M + \omega_0
\end{array}\right) 
\,. 
\label{Mfinal}
\end{equation}
The transition magnetic moment terms $\delta_0$ and $\delta_1$ 
are of higher mass dimension
and ought be much smaller in effect than $\delta$, despite the appearance of an
external magnetic field. This follows because the energy scales associated
with magnetic fields are naturally so small --- note that 
$|\mu_n| \approx 60\, \hbox{neV/T}$. 
We employ naive dimensional analysis to flesh out our assessment. 
That is, we estimate the 
$n$-${\bar n}$ matrix element associated with the leading operator, 
of mass dimension nine, as 
$\kappa \Lambda_{\rm QCD}^6/M_{n{\bar n}}^5$~\cite{Kronfeld:2013uoa}, 
where $\kappa$ is a
dimensionless constant presumably of ${\cal O}(1)$, $M_{n{\bar n}}$ is
the scale of $n$-${\bar n}$ mixing, and 
$\Lambda_{\rm QCD} \sim 200$~MeV. Writing 
$\mu_{n {\bar n}} B = (\mu_{n {\bar n}}/|\mu_n|)|\mu_n| B$, noting
$\mu_{n {\bar n}}/|\mu_n| 
\sim \kappa^\prime (\Lambda_{\rm QCD}/M_{n {\bar n}})^7$
with $\kappa^\prime$ a dimensionless constant, we estimate
$\mu_{n {\bar n}} B/\delta \sim (\kappa^\prime /\kappa)
\Lambda_{\rm QCD} |\mu_n| B / M_{n {\bar n}}^2$. 
Even in the environment of a pulsar, for which 
$B\sim 10^8$~T is possible, we see that 
$|\mu_n| B$ is many orders of magnitude smaller than 
$\Lambda_{\rm QCD}$ --- so that $\mu_{n {\bar n}} B$ is 
negligible relative to $\delta$ 
if we assume $\kappa^\prime/\kappa \sim {\cal O}(1)$.

Before closing this section we note that it is also possible 
to have a $n$-${\bar n}$
transition electric dipole moment as well, 
though this would 
certainly require an additional new physics 
mechanism to generate an appreciable effect. 
The $n$-${\bar n}$ matrix elements of 
$\psi^T \gamma_5 \sigma^{\mu\nu} C \psi F_{\mu\nu}$ 
and its Hermitian conjugate 
yields terms of the form given in Eqs.~(\ref{tran1}) and (\ref{tran2}), 
but with
$-\mathbf{B}$ replaced with $i\mathbf{E}$. 
These operators are \textbf{CP} and \textbf{T} even but \textbf{P} odd.

{\em 4. Examples.} 
In what follows we consider concrete examples of how applied magnetic fields
can be used to evade 
the quenching of $n$-${\bar n}$ oscillations found in earlier
work~\cite{Mohapatra:1980de,Cowsik:1980np}. 
We consider the leading $n$-${\bar n}$
transition operator matrix element exclusively, so that 
we rely on SM effects to realize this. 
To compute the transition probabilities, we must first 
find the normalized eigenvectors of the Hamiltonian matrix in terms
of our chosen $\{ |n+\rangle, |n-\rangle, |{\bar n}+\rangle, |{\bar n}-\rangle \}$ 
basis; we denote a state of the latter by
$| n_i \rangle$ and a normalized eigenvector by
$| u_i \rangle$ with associated 
eigenvalue $\lambda_i$, noting $i\in 1,\dots, 4$. 
The time evolution of a state of the Hamiltonian is thus given by 
\begin{equation}
| \psi(t) \rangle = \sum_{i=1}^{4} e^{-i\lambda t} 
\langle u_i | \psi (0) \rangle \, | u_i \rangle \,.
\end{equation}
Letting $|\psi(0) \rangle = | n_k \rangle$ and defining 
$a_{ij} \equiv \langle n_j | u_i \rangle$, we find 
\begin{eqnarray}
{\cal P}_{n_k \to n_j} = 
\left| \sum_{i=1}^4 e^{-i \lambda_i t} a_{ij} a^\ast_{ik} \right|^2 \,.
\end{eqnarray}
For reference, 
we find in the absence of  magnetic fields that 
${\cal P}_{n\to {\bar n}} = \sin^2 (\delta t)$, identical to that
found using Eq.~(\ref{eq2by2})~\cite{Mohapatra:1980de}. 

As a first example, we consider 
a system with a static magnetic field $\mathbf{B}_0$,
serving as the quantization axis, to
which a static transverse field $\mathbf{B}_1$ is suddenly applied at $t=0$. 
For $t > 0$ the mass matrix has the form of Eq.~(\ref{Mfinal}) 
with $\delta_0=\delta_1=0$. Noting that $|\delta| \ll |\omega_0|\,, |\omega_1|$, we
find 
that the probability of a neutron in a $s=+$
state transforming to ${\bar n}$ of fixed spin is 
\begin{eqnarray}
\!\!\!\!\!\!\!\!\!
{\cal P}_{n+\, \to {\bar n}+}(t) &=& 
\delta^2\Bigg[ \frac{\omega_1^4 t^2}{(\omega_0^2 + \omega_1^2)^2} 
\cos^2 \left(t\sqrt{\omega_0^2 + \omega_1^2}\right) + \frac{\omega_0^4}{(\omega_0^2 + \omega_1^2)^3} 
\sin^2 \left(t\sqrt{\omega_0^2 + \omega_1^2}\right)
\nonumber \\
&&+ 
\frac{\omega_0^2 \omega_1^2 t}{(\omega_0^2 + \omega_1^2)^{5/2}} 
\Bigg]  + {\cal O}(\delta^3)
 ; \,\, \label{ex1p} \\
\!\!\!\!\!\!\!\!\!
{\cal P}_{n+\, \to {\bar n}-} (t)  &=& 
\delta^2\Bigg[ \frac{\omega_1^2 t^2}{\omega_0^2 + \omega_1^2} 
- 
\frac{\omega_1^4 t^2}{(\omega_0^2 + \omega_1^2)^2} 
\cos^2 \left(t\sqrt{\omega_0^2 + \omega_1^2}\right) 
\nonumber \\
&&+ \frac{\omega_0^2 \omega_1^2}{(\omega_0^2 + \omega_1^2)^3} 
\sin^2 \left(t\sqrt{\omega_0^2 + \omega_1^2}\right)
- \frac{\omega_0^2 \omega_1^2 t}{(\omega_0^2 + \omega_1^2)^{5/2}} 
\sin \left(2t\sqrt{\omega_0^2 + \omega_1^2}\right) \Bigg] 
\nonumber \\
&&+ {\cal O}(\delta^3) \,.
\label{ex1m}
\end{eqnarray}
If $|\omega_0| \sim |\omega_1|$, we see that the last two terms of 
Eqs.~(\ref{ex1p}) and (\ref{ex1m}) are of 
${\cal O}(\delta^2/\omega_0^2)$ and ${\cal O}(t \delta^2/\omega_0)$, respectively, so that
they are indeed quenched in a magnetic field. The other terms, however, are of
${\cal O}(1)$. We note that 
${\cal P}_{n+\, \to {\bar n}-} (t)$ is larger, since 
$\omega_1^2 /(\omega_0^2 + \omega_1^2) > (\omega_1^2 /(\omega_0^2 + \omega_1^2))^2$ in this limit
--- we had anticipated this because the two states are of the same energy. 
We note that ${\cal P}_{n+\, \to {\bar n}-} (t) = {\cal P}_{n-\, \to {\bar n}+} (t)$ 
and ${\cal P}_{n+\, \to {\bar n}+} (t) = {\cal P}_{n-\, \to {\bar n}-} (t)$, so that 
the unpolarized transition probability is 
\begin{eqnarray}
{\cal P}_{n\, \to {\bar n}} (t)  &=& \delta^2\Bigg[
\frac{\omega_1^2 t^2}{\omega_0^2 + \omega_1^2} + 
\frac{\omega_0^2}{(\omega_0^2 + \omega_1^2)^2} \sin^2 (t\sqrt{\omega_0^2 + \omega_1^2})
 \nonumber \\
&+&  \frac{\omega_0^2 \omega_1^2 t}{(\omega_0^2 + \omega_1^2)^{5/2}}
\left(1 - \sin \left(2t\sqrt{\omega_0^2 + \omega_1^2}\right)\right) \Bigg] 
+ {\cal O}(\delta^3) \,,
\end{eqnarray}
--- and the first term is of ${\cal O}(1)$. For reference, 
${\cal P}_{n+\, \to n-} (t)  = (\omega_1^2 /( \omega_0^2 + \omega_1^2))
\sin(t \sqrt{\omega_0^2 + \omega_1^2}) + {\cal O}(\delta^2)$. 
The exact eigenvalues and eigenstates for $t>0$ are 
\begin{eqnarray}
E_1 &=& M_1 - \sqrt{\omega_0^2 + (\delta - \omega_1)^2} 
\,, \quad \nonumber \\
E_2 &=& M_1 + \sqrt{\omega_0^2 + (\delta - \omega_1)^2} 
\,, \quad \nonumber \\
E_3 &=& M_1 - \sqrt{\omega_0^2 + (\delta + \omega_1)^2} 
\,, \quad \nonumber \\
E_4 &=& M_1 + \sqrt{\omega_0^2 + (\delta + \omega_1)^2} 
\end{eqnarray}
and
\begin{eqnarray}
u_1 &=& \frac{1}{\sqrt{N_1}} \left\{ 1, 
\frac{(\delta -\omega_1)}{\omega_0 - \sqrt{\omega_0^2 + (\delta - \omega_1)^2}}, 
\frac{-(\delta -\omega_1)}{\omega_0 - \sqrt{\omega_0^2 + 
(\delta - \omega_1)^2}}, 1 \right\} 
\,, \nonumber \\
u_2 &=& \frac{1}{\sqrt{N_2}} \left\{ 1, 
\frac{(\delta -\omega_1)}{\omega_0 + \sqrt{\omega_0^2 + (\delta - \omega_1)^2}}, 
\frac{-(\delta -\omega_1)}{\omega_0 + \sqrt{\omega_0^2 
+ (\delta - \omega_1)^2}}, 1 \right\} \,,
\nonumber \\
u_3 &=& \frac{1}{\sqrt{N_3}} \left\{ -1, 
\frac{-(\delta +\omega_1)}{\omega_0 - \sqrt{\omega_0^2 + (\delta + \omega_1)^2}}, 
\frac{-(\delta + \omega_1)}{\omega_0 - \sqrt{\omega_0^2 + (\delta + \omega_1)^2}}, 1 
\right\} \,, \nonumber \\
u_4 &=& \frac{1}{\sqrt{N_4}} \left\{ -1, 
\frac{-(\delta +\omega_1)}{\omega_0 + \sqrt{\omega_0^2 + (\delta + \omega_1)^2}}, 
\frac{-(\delta +\omega_1)}{\omega_0 + \sqrt{\omega_0^2 + (\delta + \omega_1)^2}}, 1 \right\} \,,
\end{eqnarray}
with 
\begin{eqnarray}
N_{\stackrel{1}{{}_{2}}} &=& 2 \left[1 + \frac{(\delta -\omega_1)^2}{\omega_0 \mp \sqrt{\omega_0^2 + 
(\delta - \omega_1)^2}}\right] \,,\nonumber \\
N_{\stackrel{3}{{}_{4}}} &=& 2 \left[1 + \frac{(\delta +\omega_1)^2}{\omega_0 \mp \sqrt{\omega_0^2 + 
(\delta + \omega_1)^2}}\right] \,.
\end{eqnarray}
If $\delta=0$ {\it or} $\omega_0=\omega_1=0$, we 
see that $E_1=E_3$ and $E_2=E_4$. In the former case, 
$u_1 + u_3$ and $u_2 + u_4$ yield linear combinations of ${\bar n}(+)$
and ${\bar n}(-)$, and $u_1 - u_3$ and $u_2 - u_4$ 
yield linear combinations of $n(+)$ and $n(-)$. In contrast, in the latter
case, we find Majorana states; that is, 
$u_1 \pm u_3\propto \Psi_M^\pm (\mp)$ and 
$u_2 \pm u_4\propto \Psi_M^\mp (\mp)$. 

As long known, 
the spin of a macroscopic sample of fermions can be made to flip 
through the use of magnetic resonance techniques. 
Indeed, supposing the spins are aligned 
(or anti-aligned) with a static magnetic
field, and an oscillatory magnetic field is applied transverse to it, 
 we can tune the frequency of the transverse field 
in such a way that the probability of flipping the neutron spin is of
${\cal O}(1)$ irrespective of the size of the applied magnetic fields 
--- this is the famous Rabi formula~\cite{PhysRev.51.652,CTDL}. 
Thus as a second example we study $n$-${\bar n}$ oscillations 
in such a magnetic
field arrangement~\cite{CTDL}, 
replacing  $B_1$ with a time-dependent magnetic field $B_1 (t)$, 
so that the SM Hamiltonian for a neutron becomes 
$H(t) = \omega_0 \,\sigma_z + \omega_1 (\cos \omega t \,\sigma_x 
+ \sin \omega t \,\sigma_y)$. 
The resulting $n$-${\bar n}$ Hamiltonian matrix is of form 
\begin{equation} 
{\cal H}(t) = \left(\begin{array}{cccc}
M + \omega_0 & \delta & \omega_1 e^{- i \omega t} & 0  \\
\delta    &  M  -\omega_0 & 
0 & -\omega_1 e^{- i \omega t}  \\
\omega_1 e^{ i \omega t}   & 0 
 & M -\omega_0 & - \delta 
\\
0 
& -\omega_1 e^{ i \omega t}  
  & -\delta   & M + \omega_0
\end{array}\right) 
\,. 
\label{Mrffinal}
\end{equation}
To compute the transition probabilities in this case, 
we solve the time-dependent Schr\"odinger equation 
$i \partial_t \psi = {\cal H} \psi$ with 
$\psi= \{a_+ (t), {\bar a}_+ (t), a_- (t), {\bar a}_- (t)\}$ 
through the change of variable
$\stackrel{{}_{(-)}}{a_\pm} = \stackrel{{}_{(-)}}{b_\pm} 
\exp(\mp i\omega t/2)$. This yields 
$i \partial_t {\tilde \psi} = {\tilde {\cal H}}{\tilde \psi}$ with 
${\tilde \psi} = \{ b_+ (t), {\bar b}_+ (t), b_- (t), {\bar b}_- (t) \}$ 
and 
\begin{equation} 
{\tilde {\cal H}} = \left(\begin{array}{cccc}
M - \Delta\omega_- & \delta & \omega_1  & 0  \\
\delta    &  M  - \Delta\omega_+ & 
0 & -\omega_1  \\
\omega_1    & 0 
 & M + \Delta\omega_- & - \delta 
\\
0 
& -\omega_1 
  & -\delta   & M + \Delta\omega_+
\end{array}\right) 
\, 
\label{Mrfconv}
\end{equation}
with $\Delta\omega_{\pm}\equiv \omega/2 \pm \omega_0$, noting 
that the transition probabilities of 
interest follow immediately from its solution 
because $|\stackrel{{}_{(-)}}{a_\pm}|^2 = |\stackrel{{}_{(-)}}{b_\pm}|^2$. 
The oscillatory transverse field needed for 
magnetic resonance experiments is typically 
realized, however, through the application of a
radio frequency (rf) field with linear polarization, so that 
if $\Delta \omega_+=0$, then $\Delta \omega_-=0$ also. 
Thus under usual experimental conditions the largest contributions have 
$\Delta \omega_+  = - \Delta \omega_{-}$, and the 
$n$-${\bar n}$ transition probabilities can be estimated from 
Eqs.~(\ref{ex1p}) and (\ref{ex1m}) upon the 
replacement $\omega_0 \rightarrow \Delta \omega_+$. 
On resonance, for which $\Delta \omega_\pm =0$, 
we have 
\begin{eqnarray}
\!\!\!\!\!\!\!\!\!
{\cal P}_{n+\, \to {\bar n}+}(t) &\approx& 
\delta^2 t^2 
\cos^2 \left(t\sqrt{\omega_0^2 + \omega_1^2}\right) + {\cal O}(\delta^3)
 ; \,\, \label{ex1pres} \\
\!\!\!\!\!\!\!\!\!
{\cal P}_{n+\, \to {\bar n}-} (t)  &\approx& 
\delta^2 t^2 
\sin^2 \left(t\sqrt{\omega_0^2 + \omega_1^2}\right) + {\cal O}(\delta^3) \,, 
\label{ex1mres}
\end{eqnarray}
where we have neglected contributions controlled by 
$|\omega|/2 + \omega_0$ as per standard practice~\cite{Ramsey}.
Finally, we find, similarly, that 
the unpolarized transition probability is 
${\cal P}_{n\, \to {\bar n}}(t) \approx  \delta^2 t^2 + {\cal O}(\delta^3)$. 

{\em 6. New Experimental Prospects.} 
We have shown through explicit example that the removal of 
magnetic fields is not  necessary for the observation of $n$-${\bar n}$ 
oscillations; this opens new possibilities for their experimental 
discovery. 
For example, it becomes possible to 
study $n$-${\bar n}$ oscillations by confining neutrons in 
magnetic traps, or bottles; such are under 
development for improved measurements of the neutron 
lifetime~\cite{Leung:2008af,Walstrom:2009,Salvat:2013gpa}. 
In a gravitomagnetic trap a single spin state is confined; we
suppose, in addition, that 
a transverse rf field at resonance is applied. 
If the spin-flip time is short  
compared to the time for a confined neutron to be lost from the trap, 
we suppose that the storage time determined 
under these conditions can 
be used to set a limit on $n$-${\bar n}$ oscillations. 
That is, an experimental limit on $n$-${\bar n}$ oscillations 
can be defined by writing the 
transition probability as
${\cal P}_{n\to {\bar n}} \simeq (t/\tau_{n{\bar n}})^2$
and bounding $\tau_{n {\bar n}}$. 
A crude estimate of the oscillation lifetime is given by
$(\tau_{n\bar n})_{\rm bottle} \sim 
\left( N_{\rm fill} N_{\rm trial} \langle t^2 \rangle/ {\bar N} \right)^{1/2}$, 
where $N_{\rm fill}$ is the number of neutrons (i.e., $nV$ with $n$ the neutron
number density and $V$ the volume of the trap) added to the bottle
at one time, $N_{\rm trial}$ is the number of times the trap is filled, 
${\bar N}$ is the limit on the number of antineutrons detected, 
and
$\langle t^2 \rangle^{1/2}$ 
is the storage time in the trap. Estimating
$N_{\rm fill}\sim 10^7$, 
$N_{\rm trial}\sim 10^5$, and 
$\langle t^2 \rangle^{1/2} \sim 400 \,{\rm s}$
and using ${\bar N} \le 2.3$ at 90\% C.L.~\cite{BaldoCeolin:1994jz}
 yields 
$\tau_{n {\bar n}} \sim 2\times 10^8\,{\rm s}$, so that the gain 
seems modest over the existing limit of 
$\tau_{n {\bar n}} \ge 0.86 \times 10^{8} \,{\rm s}$ at 
90\% C.L.~\cite{BaldoCeolin:1994jz}, 
though one can expect further improvements with bettered 
ultracold neutron sources. 

%%%%%%%%%%%%%%%%%%%%%%%%%%%%%%%%%%%%%%%%%%%%%%%%%%%%%%%%%%%%%%%%%%%%%%%%%%%%%%%
{\em 7. Summary.} 
As long recognized, the discovery of ${\cal B}-{\cal L}$ 
violation would speak to the existence of Majorana dynamics
in Nature. This would not imply, however, that the neutron is its own antiparticle, 
but, rather, that the weak interaction eigenstates of the $n$-${\bar n}$ 
system in vacuum 
transform into themselves under the charge conjugation operator $\mathbf{C}$. 
Although many 
authors~\cite{Arndt:1981ey,Trower:1982qs,Pusch:1982ps,Krstic:1988ix,Dubbers:1989pa} 
have studied the impact of external
magnetic fields on $n$-${\bar n}$ oscillations 
within the context of the $2\times 2$ 
phenomenological framework~\cite{Mohapatra:1980de}, 
our work is the first 
to incorporate spin in a fundamental way. The results 
that emerge are remarkably different from earlier studies --- 
in particular, magnetic field mitigation is not required to observe
$n$-${\bar n}$ mixing, as had been previously thought~\cite{Kronfeld:2013uoa,ESS}.

%%%%%%%%%%%%%%%%%%%%%%%%%%%%%%%%%%%%%%%%%%%%%%%%%%%%%%%%%%%%%%%%%%%%%%%%%%%%%%%
\begin{acknowledgments}
We thank S. Ramachrandan for 
contributing to an early phase of this work, and 
we acknowledge partial support by the Department
of Energy Office of Nuclear Physics under 
contract DE-FG02-96ER40989. S.G. thanks 
G. Greene for keen interest and useful suggestions, 
N. Brambilla and the
Excellence Cluster of the Technical University of Munich
for gracious hospitality, 
and Y. Kamyshkov for references. We also thank S. Syritsyn for comments
on an earlier version of this paper.

\end{acknowledgments}
%%%%%%%%%%%%%%%%%%%%%%%%%%%%%%%%%%%%%%%%%%%%%%%%%%%%%%%%%%%%%%%%%%%%%%%%%%%%%%%
%%%%%%%%%%%%%%%%%%%%%%%%%%%%%%%%%%%%%%%%%%%%%%%%%%%%%%%%%%%%%%%%%%%%%%%%%%%%%%%
%\bibliographystyle{doiplain}
\bibliography{revnnbar}

\end{document}